\documentclass{aa}
\usepackage{latexsym}
\usepackage{natbib}
\usepackage{graphicx}
\newcounter{IonCS}
\renewcommand{\ion}[2]{\setcounter{IonCS}{#2}#1\,{\scshape{\roman{IonCS}}}}
\newcommand{\sect}[1]{Sect.\,\ref{#1}}
\newcommand{\sects}[1]{Sects.\,\ref{#1}}
\newcommand{\fig}[1]{Fig.\,\ref{#1}}
\newcommand{\figs}[1]{Figs.\,\ref{#1}}

\newcommand{\unit}[1]{{\rm{#1}}}

\graphicspath{{./}{figs/}}

\def\singlefigwid{8.8cm}

\begin{document}

%
\title{Catastrophic cooling and cessation of heating in the solar corona}

\titlerunning{Catastrophic cooling and cessation of heating in the solar corona}

\authorrunning{H. Peter et al.}

\author{H.~Peter, S.~Bingert, S.~Kamio}

\institute{Max-Planck-Institut f{\"u}r Sonnensystemforschung, 
           37191 Katlenburg-Lindau, Germany, email: peter@mps.mpg.de
           }

\date{Received 16 August 2011 / Accepted 12 November 2011}

\abstract%
%
{%
Condensations in the more than $10^6$\,K hot corona of the Sun are commonly observed in the extreme ultraviolet (EUV). 
While their contribution to the total solar EUV radiation is still a matter of debate, these condensations certainly provide a valuable tool for studying the dynamic response of the corona to the heating processes.
}
{%
We investigate different distributions of energy input in time and space to investigate which process is most relevant for understanding these coronal condensations.
}
{%
For a comparison to observations we synthesize EUV emission from a time-dependent, one-dimensional model for coronal loops, where we employ two heating scenarios: simply shutting down the heating and a model where the heating is very concentrated at the loop footpoints, while keeping the total heat input constant.
}
{%
The heating off/on model does not lead to significant EUV count rates that one observes with SDO/AIA. In contrast, the concentration of the heating near the footpoints leads to thermal non-equilibrium near the loop top resulting in the well-known catastrophic cooling. This process gives a good match to observations of coronal condensations.
}
{%
This shows that the corona needs a steady supply of energy to support the coronal plasma, even during coronal condensations. Otherwise the corona would drain very fast, too fast to even form a condensation.
}
%

\maketitle

\section{Introduction\label{S:intro}}

Loops in the corona still present us with numerous open questions. Besides the open fundamental problem of the coronal heating mechanism, it is not clear yet which processes can provide the high densities observed high in the corona, well above what would be expected by hydrostatic equilibrium \cite[e.g.][]{Peres:1997,Aschwanden+al:2001}.
One candidate for this is the thermal non-equilibrium that leads to \emph{catastrophic cooling}. This process is initiated if the heating is very concentrated at the loop footpoints, and the heating at the loop top can no longer balance the losses from heat conduction and radiation: the loops starts cooling, radiation becomes more efficient, and a runaway process sets in \citep{Kuin+Martens:1982}. This process is considered to be important in the formation of prominences \citep{Antiochos+Klimchuk:1991,Karpen+al:2006}.
The role of this \emph{catastrophic cooling} for normal coronal loops has been studied in several numerical experiments that employ one-dimensional models for (semi-circular) loops. \cite{Mueller+al:2003,Mueller+al:2004} have investigated the cyclic and chaotic behaviour for loops of different lengths and derived observable quantities such as extreme ultraviolet emission line spectra. They relate these events to ``coronal rain'' \citep{Mueller+al:2005}. \cite{Antolin+al:2010} proposes that ``coronal rain''  could be a marker of the coronal heating mechanism.

In a recent study \cite{Klimchuk+al:2010} have been able to show that possibly the catastrophic cooling associated with the thermal non-equilibrium does not play a major role in active region coronal loops.
Consequently, \cite{Klimchuk+al:2010} argue that the heat input for active region loops is probably not concentrated near the loop's footpoints alone. The authors furthermore suggest that the heating is probably not steady.

The conclusions of \cite{Klimchuk+al:2010} on the relevance of the process of thermal non-equilibrium are still under debate, as has become clear in discussions at the recent Loops-
5 workhop (June 2011).  When the loop cross-section is allowed to vary, and especially when the heating profile is made asymmetric, the nature of catastrophic cooling may be significantly different than concluded by \cite{Klimchuk+al:2010}.  It may be premature
to rule out thermal non-equilibrium for explaining some of the observed properties of coronal loops (Z. Miki\'c, priv. comm.).

Three-dimensional magneto-hydrodynamic models (3D MHD) of the corona show a horizontally \emph{averaged} exponential drop in the heating rate with altitude, with a scale height of the order of 5\,Mm \citep[e.g.][]{Gudiksen+Nordlund:2002,Gudiksen+Nordlund:2005a,Gudiksen+Nordlund:2005b}, and the heating is found to be intermittent in time and space \citep{Bingert+Peter:2011}. The assumptions of the \cite{Klimchuk+al:2010} study concerning the spatial and temporal distribution of the heat input are consistent with the \emph{average} drop of the heating rate found in the 3D MHD models.
This type of 3D MHD model can be considered realistic, since they successfully reproduce numerous properties observed in imaging and spectroscopy, such as the emission measure distribution or the average transition region Doppler shifts \citep[e.g.][]{Peter+al:2004,Peter+al:2006,Zacharias+al:2011.doppler}.
Because these 3D models predict a heating scale height greater than what is needed for catastrophic cooling, this is further evidence that \emph{extreme} concentration of the heating rate towards the footpoints is not a \emph{general} feature of coronal heating.

One can consider catastrophic cooling as a process that will happen only in a sub-volume of the corona, where the heating (for some time) is concentrated much more towards the loop footpoints than on average. These catastrophic cooling events are observed and seem to be a common feature. Examples of detailed observational studies of such events are the investigations by \cite{Schrijver:2001:cooling} in a comparison of filtergrams in Lyman-$\alpha$ and in the coronal band around 171\,{\AA}, by \cite{deGroof+al:2005} when comparing the 171\,{\AA} band with H-$\alpha$ data, 
or by \cite{Antolin+Verwichte:2011} on the role of these condensations for loop oscillations.

Most recently, cooling events have been reported by \citet{Kamio+al:2011.upflow} using data from the Atmospheric Image Assembly on the Solar Dynamics Observatory \citep[AIA/SDO,][]{Lemen+al:2011}.
In particular, \citet{Kamio+al:2011.upflow} present cases with peculiar light curves: the 131\,\AA\ and 171\,\AA\ representing plasma at ${\log}T\,[\mbox{K}]{\approx}5.7$ and 5.9 show narrow single peaks in time. In contrast, the 193 \,\AA\ and 211\,\AA\ channels with a maximum contribution for plasma at ${\log}T\,[\mbox{K}]{\approx}6.2$ and 6.3 show enhanced emission for much longer times with a more complex light curve, including peaks before and after the peaks in the ``cooler'' channels.
A more detailed presentation of this lightcurve and the comparison to our model can be found in \sect{S:obs}.
The lightcurve found by \citet{Kamio+al:2011.upflow} is counter-intuitive for a situation where the plasma cools, and one would expect a brightening in the channels according to their temperature of maximum contribution. A simplistic model was presented by \citet{Kamio+al:2011.upflow} to give a first attempt to explain this peculiar behaviour assuming a cooling at constant pressure.

In the present study we investigate the cooling of the plasma in order to understand the peculiar light curves found by \citet{Kamio+al:2011.upflow}. We perform one-dimensional loop models starting from an equilibrium model that is hot enough that it does not show significant count rates in any of the above channels. We  then follow two scenarios: (1) a \emph{catastrophic cooling} case where the heating is concentrated at the foot points, and (2) a \emph{heating off/on} case where the heating is shut off completely for a while and the turned on again. In both cases the plasma will cool, but only the first case will be compatible with the observed cooling events.

After a short outline of the model in \sect{S:model} we present the hydrodynamic results (\sect{S:results}) and the observable AIA signatures (\sect{S:signature}) for both cases, before we discuss a comparison to the observations of \citet{Kamio+al:2011.upflow} in more detail in \sect{S:obs} and conclude the paper.

\section{1D loop model and synthesizing observations\label{S:model}}

To model the dynamics of a coronal loop we solve the mass, momentum, and energy balance in a 1D model for a semi-circular loop with constant cross-section. For the numerical experiments we employ the Pencil code \citep[][]{Brandenburg+Dobler:2002}%
\footnote{%
pencil-code.googlecode.com/
} 
with modifications to account for the physics of the corona \citep{Bingert+Peter:2011}. The equations (in their full 3D form) and a brief description of the method can be found in \cite{Bingert+Peter:2011}.
 
For the loop model we assume that all variables are a function of the coordinate $s$ along a magnetic field line, which is assumed to be semi-circular. The velocity is only along the loop. At both ends we impose boundary conditions reflecting the photosphere; i.e., we prescribe the density and the temperature. The velocity is set to zero at both ends.
In the 3D-model of \cite{Bingert+Peter:2011}, the heating through Ohmic dissipation is self-consistently described and results in a roughly exponential decrease in the heating rate in the coronal part with a scale height of about 5\,Mm. In this 1D model we simply define the heating rate to be exponentially decreasing and constant in time (see \sect{S:heating}). The energy equation also accounts for optically thin radiative losses \citep[from][]{Cook+al:1989} and heat conduction \citep{Spitzer:1962}, which is essential to properly model the thermal non-equilibrium leading to catastrophic cooling. In the momentum equation, gravity is taken into account.

The model produces a hot corona above a cool photosphere and chromosphere at both footpoints with a thin transition region. The lower dense and cool part of the loop is only used as a reservoir for mass and energy, so we do not solve the radiative transfer problem, among others. For the simulations shown in this manuscript we used 2048 gridpoints along the loop.

\subsection{Heating rate and equilibrium model\label{S:heating}}

For the heating rate we prescribe an exponentially decreasing volumetric heating rate,
\begin{equation}\label{E:heating.density}
Q =  Q_{\rm{0}}\,\exp\left(-\frac{z}{\lambda}\right)~,
\end{equation}
with the heating rate at the lower boundary $Q_0$ and the scale length of the heating rate $\lambda$. This is a function of the height $z$ and not of the arc length along the loop. The energy flux density into the corona $F_{\rm{H}}$ is the flux at the base of the corona,
\begin{equation}\label{E:heating.flux}
F_{\rm{H}}=\int_{z_{\rm{base}}}^{z_{\rm{top}}} Q\,{\rm{d}}z ~,
\end{equation}
where $z_{\rm{top}}$ is the height of the loop apex. We define the base of corona, $z_{\rm{base}}$, as the height where the transition region starts, i.e., where a temperature of $10^4$\,K is reached. Essentially, there are two free parameters for the heating rate, namely the energy flux into the corona, $F_{\rm{H}}$, and the degree of concentration of the heating towards the footpoints parameterized through the heating scale length $\lambda$.

To obtain a thermal non-equilibrium causing a catastrophic cooling event, the heating scale length $\lambda$ has to be short enough compared to the loop length. Because we want to compare an event with another one without catastrophic cooling, we choose $\lambda{=}2$\,Mm. This provides a stable solution (while $\lambda{=}1$\,Mm shows a loss of equilibrium) for the loops with a length of $L{=}120$\,Mm that we study here. This loop length is motivated by the observation of \cite{Kamio+al:2011.upflow}, who find the condensation at an apex height of about 40\,Mm.

Because we intend to investigate a loop that is initially too hot to be seen in the SDO/AIA 193\,{\AA} or 211\,{\AA} channels, the apex temperature should be at least ${\log}\,\widehat{T}\,\unit{[K]\approx6.5}$. This automatically sets the required energy flux to heat the corona. 
In our equilibrium (numerical) model we choose a heating rate resulting in a coronal base at $z_{\rm{base}}\approx3.4$\,Mm with a rather high coronal energy input of $F_{\rm{H}}\approx15\,000\,\unit{W\,m^{-2}}$. This gives the required apex temperature of ${\log}\,\widehat{T}\,\unit{[K]\approx6.5}$. This choice of parameters automatically sets the pressure in the loop and thus the density at the apex, which is about ${\log}\,\widehat{n}\,\unit{[cm^{-3}]\approx9.5}$ in the equilibrium model.

The heat input (and the loop length) fully determines the temperature \emph{and} the density. This is also reflected by the scaling laws as derived by \cite{Rosner+al:1978},
\begin{eqnarray}
\label{E:RTV.T}
\widehat{T}\,\unit{[K]} & = & 1700 ~ \big( F_{\rm{H}}\,\unit{[W\,m^{-2}]} \big)^{2/7}
                                    ~ \big( L\,\unit{[m]} \big)^{2/7}~,
\\
\label{E:RTV.n}
\widehat{n}\,\unit{[cm^{-3}]} &=&
                  3.9{\cdot}10^{10} ~ \big( F_{\rm{H}}\,\unit{[W\,m^{-2}]} \big)^{4/7}
                                    ~ \big( L\,\unit{[m]} \big)^{-3/7}~.
\end{eqnarray}
In this modified form the (constant) heating rate $E_{\rm{H}}$ of \cite{Rosner+al:1978} is translated into the energy flux density into the corona, $F_{\rm{H}}=E_{\rm{H}}\,(L/2)$. For the above values of energy flux and loop length used for our equilibrium model, these scaling laws give comparable values for the temperature and density at the loop apex as the numerical model, as should be expected.

\subsection{Numerical experiments\label{S:experiments}}

We started the two different numerical experiments from an equilibrium model with the parameters for the heating rate as mentioned in \sect{S:heating}. In particular, the heating scale length is $\lambda=2$\,Mm. We let this model run for a long time to be sure that it represents an equilibrium solution. ``Long'' means much longer than the coronal cooling time, which is around\,hour.

The first systematic study for loops heated with the exponentially decreasing heating rate used here was performed by \cite{Serio+al:1981}. For small loops (shorter than the pressure scale height), they could find solutions for their \emph{static} model with a temperature peak at the apex only if the heating scale height was more than half the loop length. For longer loops they found this threshold is a third of the pressure scale height. Thus this would be about 50\,Mm in our case (of long loops), and according to \cite{Serio+al:1981}  we should not find a stable solution for $\lambda=2$\,Mm. However, later \emph{time-dependent} models such as the ones by \cite{Mueller+al:2003,Mueller+al:2004} have found stable solutions, also at lower values for $\lambda$. Whether the thermal non-equilibrium sets in depends on the balance of heating and radiative losses (at the loop apex). This is a complex non-linear balance, because decreasing the scale length, and thus decreasing the heating at the apex, also leads to a lower density at the apex, hence less radiation. Therefore the quantitative results for the threshold value of $\lambda$ based on the static models of \cite{Serio+al:1981} do not hold in general.
Furthermore, for the present study it is not so important on which heating scale height the thermal non-equilibrium sets in, but only that it sets in at all.

\subsubsection{Catastrophic cooling}\label{S:setup.catastrophic}

In one experiment we decrease the heating scale length to $\lambda=1$, while keeping the energy flux into the corona, $F_{\rm{H}}$ as defined in (\ref{E:heating.flux}), constant.
Thus we changed $Q_0$ in (\ref{E:heating.density}) accordingly. All other parameters are kept exactly the same.

This setup will result in a thermal non-equilibrium near the top of the loop, leading to catastrophic cooling and the formation of a condensation. After a while the loop recovers its former temperature (and density). This happens in a similar fashion to the one described by \cite{Mueller+al:2003,Mueller+al:2004}.

\subsubsection{Heating off/on}\label{S:setup.normal}

In the other experiment, we kept the same heating scale height, but practically shut off the heating by decreasing the heat input by many orders of magnitude ($10^{12}$) from one time step in the simulation to the next. All other parameters were kept as in the equilibrium model. Starting with the equilibrium solution this implies that the loop cools down and the plasma will drain.

After some time we resumed the original heating rate, and the loop found its former equilibrium. We waited before turning the heating on for a time comparable to the time it takes the catastrophically cooling loop (\sect{S:setup.catastrophic}) to recover from its cooling event.

\begin{figure}
\includegraphics[width=\singlefigwid]{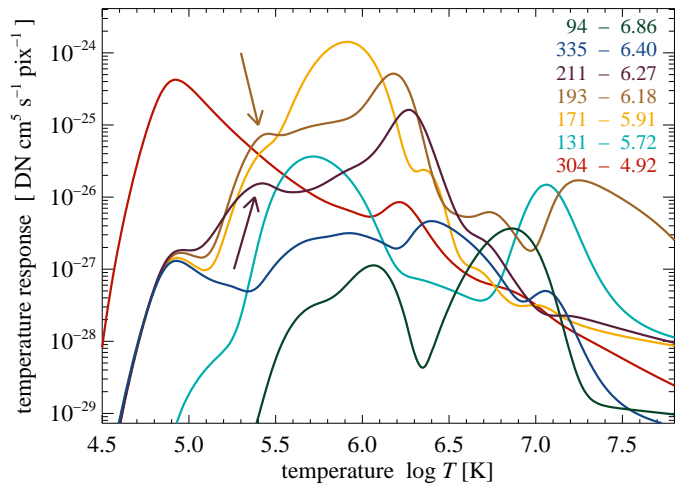}
\caption{Temperature response functions of the AIA channels as stored in SolarSoft at the time of publication of this study; following \cite{Boerner+al:2011}.
Noted are the centre wavelengths in \AA\ and the peak of the temperature response in ${\log}T[\mbox{K}]$. The secondary peaks at low temperatures in the 193\,\AA\ and 211\,\AA\ channels are indicated by arrows.
See \sect{S:synthesis}.
\label{F:response}}
\end{figure}

\subsection{AIA synthetic data\label{S:synthesis}}

To allow for a comparison with SDO/AIA observations, we synthesized the count rates to be expected if AIA were to observe the structure we model. For this we use the temperature response functions as provided in the SolarSoft package\footnote{http://www.lmsal.com/solarsoft/}. The temperature response function or kernel $K_i(T)$ is basically the integral of the contribution function of the lines in the respective bandpass $i$ weighted with the effective area as a function of wavelength. The kernels $K_i(T)$ are described in detail in \cite{Boerner+al:2011} and are displayed in \fig{F:response}.
\cite{Boerner+al:2011} use different abundances and another ionization equilibrium than \cite{Cook+al:1989} used to derive their radiative loss function, which we employ in our model. This inconsistency should not significantly affect our results, but should be checked in the future.

The emissivity (i.e., radiated power per volume) at each gridpoint of the loop model is given by $n^2\,K(T)$, with $n$ and $T$  the number density and temperature at that gridpoint. To derive the AIA counts in the common digital number DN\,pixel$^{-1}$\,s$^{-1}$, one then has to integrate along the line of sight. Because the loop model is one-dimensional, we have to assign a diameter to the loop to derive the synthetic count rate. For the investigations presented here we assume a diameter of 50 km. This choice is motivated by the diameter of the individual loop strands in the multi-stranded loop model of \cite{Patsourakos+Klimchuk:2006}, but the results presented here do not depend much on this special choice.

\begin{figure*}
\centerline{\includegraphics{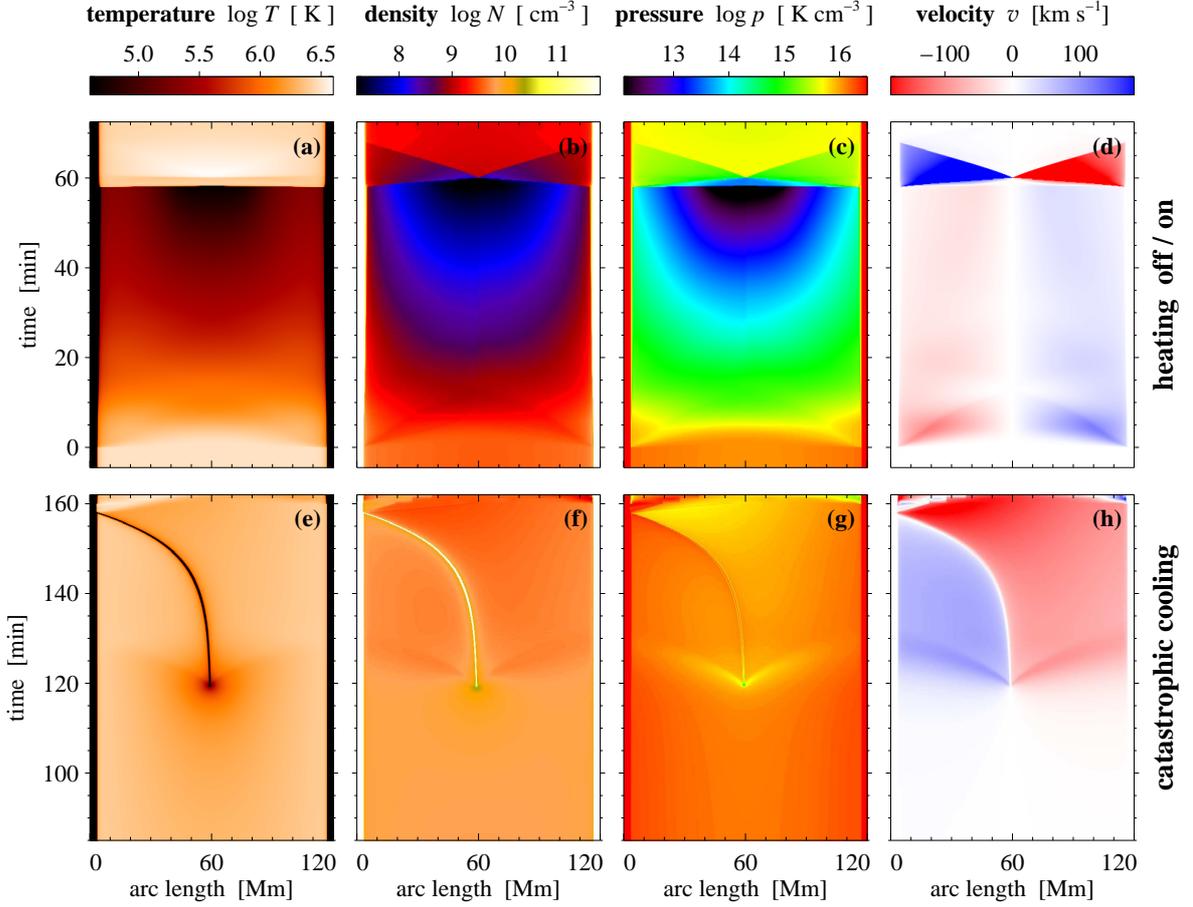}}
\caption{Plasma parameters in numerical loop models as a function of arc length along loop and time.
The loop is semi-circular with the apex at 60\,Mm and the footpoints at 0 and 120\,Mm. Shown are temperature, density, pressure, and velocity (from left to right) for the two numerical experiments.
%
The top row displays the results where the heating was shut off (at $t{=}0$) and then  turned on again at $t{\approx}57$\,min (see \sect{S:results.cooling}). 
In the bottom row, the results for the case of catastrophic cooling are shown (see \sect{S:results.catastrophic}), where $t{=}0$ refers to the time when the heating scale length was reduced. The condensation sets in at around $t{\approx}120$\,min.
In both cases the time axis spans about 70\,min.
Positive velocities (shown in blue) are in the direction of increasing arc length.
%
\label{F:temporal.global}}
\end{figure*}

\section{Results for 1D loop models\label{S:results}}

\subsection{Loop undergoing catastrophic cooling\label{S:results.catastrophic}}

\begin{figure*}
\includegraphics[width=\singlefigwid]{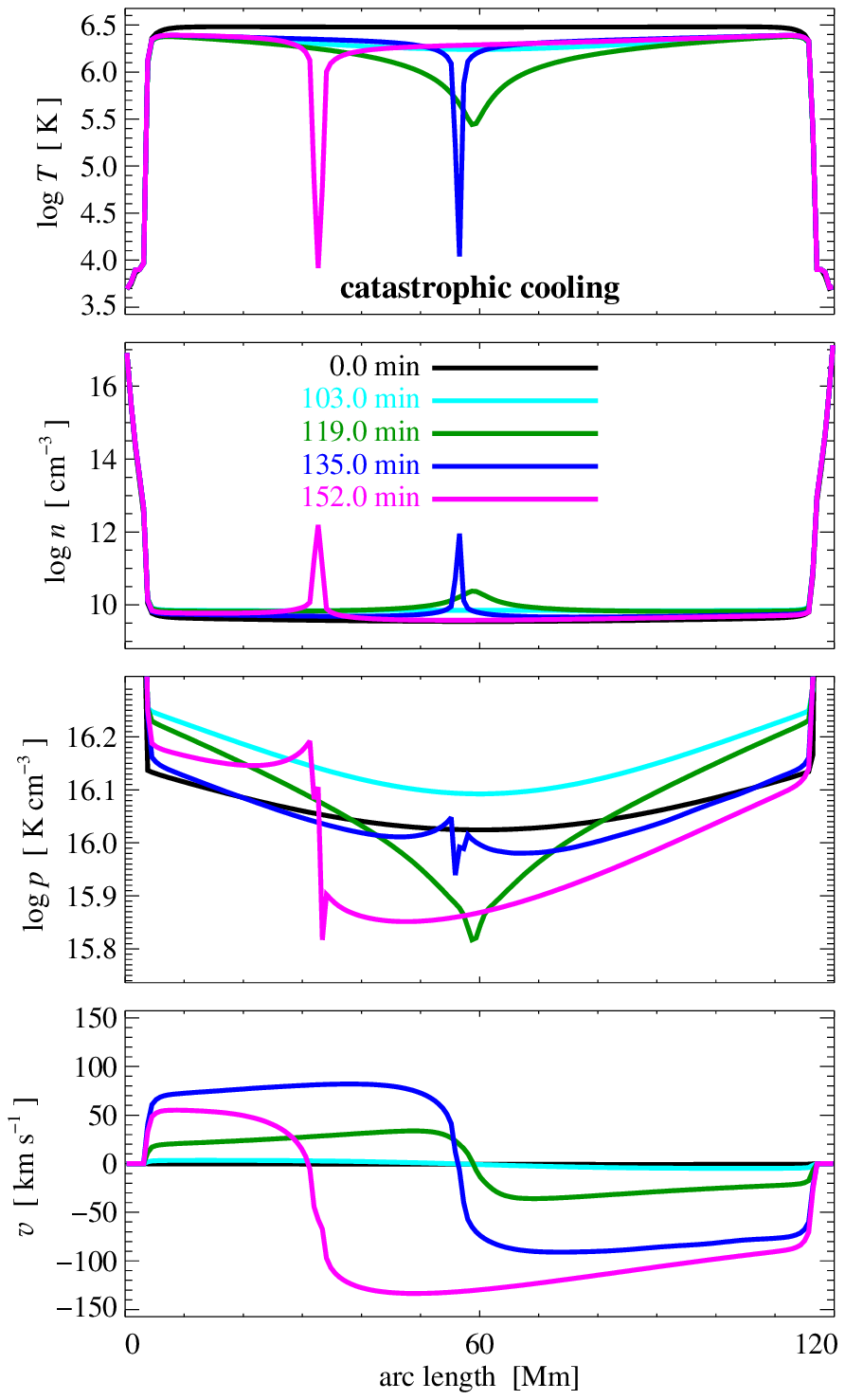} \hfill
\includegraphics[width=\singlefigwid]{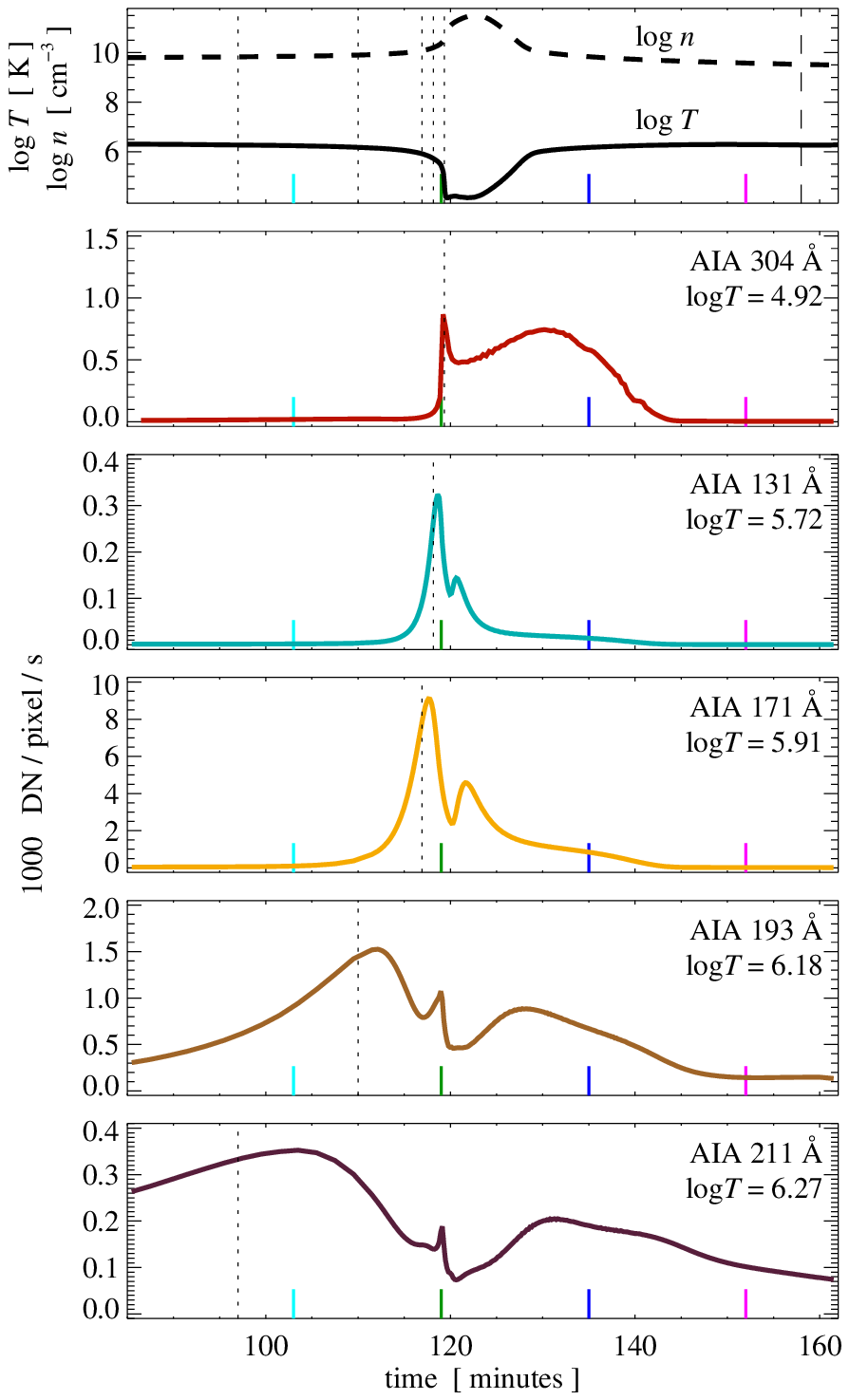}
\caption{Plasma properties and synthesized emission of AIA channels for 1D loop model undergoing catastrophic cooling.
\emph{The left column} shows snapshots of the profiles of temperature, ${\log}\,T$, density, ${\log}\,n$, pressure ${\log}\,n$, and velocity, $v$ along the loop ($v{>}0$ in direction of positive arc length).
\emph{The right column} shows the temporal evolution of the temperature and density at the loop apex (top right). The other panels in the right column show the expected count rates for several AIA/SDO channels if the loop top is observed edge on (one AIA pixel covering apex at 38\,Mm height).
The short vertical markers indicate the time for which the snapshots are shown in the left column.
The vertical dotted lines indicate the times when the temperature at the loop top (top panel) reaches the temperature of maximum response for the AIA channels. These temperatures are given (in ${\log}T$\,[K]) with each panel.
See \sects{S:results.catastrophic} and \ref{S:signature.catastrophic}.
A movie of the temporal evolution shown in the left panels is available in the on-line edition.
\label{F:condensation}}
\end{figure*}

Starting from the equilibrium model, the loop gradually evolves after having decreased the scale length $\lambda$ of the heating rate (cf.\ \sect{S:setup.catastrophic}). For approximately the first 1.5 hours the changes in the plasma parameters, i.e., temperature, pressure, and velocity are only very small. Therefore, in \fig{F:temporal.global} (bottom row) we show the evolution of these quantities for this model only after $t{\approx}90$\,min. In this early phase the loop reaches a high temperature with a flat peak at about ${\log}T\,[K]=6.5$ (see \fig{F:condensation}; top left). Initially the decrease in the heating rate puts the loop only slightly out of equilibrium. Because the initial peak temperature is high, the (radiative) cooling time is long, and thus the slow early evolution is to be expected. This is consistent with earlier work by \cite{Mueller+al:2003,Mueller+al:2004}, who find that the time between the condensations range from one to several hours, depending on loop length and heating scale length. Depending on its density and its density contrast to the surrounding corona, the initial loop might be visible in observations either as an individual X-ray coronal loop (stable for well above an hour) or as part of the diffuse background corona.

Finally, at about $t{=}110$\,min the loss of equilibrium becomes increasingly faster and a cool dense condensation forms at about $t{=}120$\,min. While cooling down, the pressure across the condensation is roughly constant, which is due to the inflow of plasma from both sides along the loop leading to an increase in density. This behaviour can be found in a quantitative fashion through the profiles shown in the left-hand panel of \fig{F:condensation}.
It is also clearly evident in the space-time plot in the bottom row of \fig{F:temporal.global}, where one sees the condensation in the temperature plot, but only a comparably weak trace in the pressure plot. Of course, there is some change in the pressure during the condensation, about 0.2 in $\log_{10}$, equivalent to a factor of 1.5. However, this is small compared to the change in temperature (or density), which is almost 2.5 in $\log_{10}$, equivalent to a factor of 300. Therefore, the catastrophic cooling can be considered to be more or less isobaric.

After hovering for some time close to the apex, the condensation slides down the loop and hits the photosphere. In this model run the condensation happens to be located just a bit left of the loop top (when looking at \fig{F:temporal.global}), and thus the condensation slid to that side. This is caused by slight asymmetries, e.g., by waves travelling back and forth through the loop following the change in the heating scale length at $t{=}0$. All of this behaviour of the condensation is consistent with previous work on catastrophic cooling in loops \citep[e.g.][]{Mueller+al:2003,Mueller+al:2004}.

\subsection{Loop with heating shut off and on again\label{S:results.cooling}}

\begin{figure*}
\includegraphics[width=\singlefigwid]{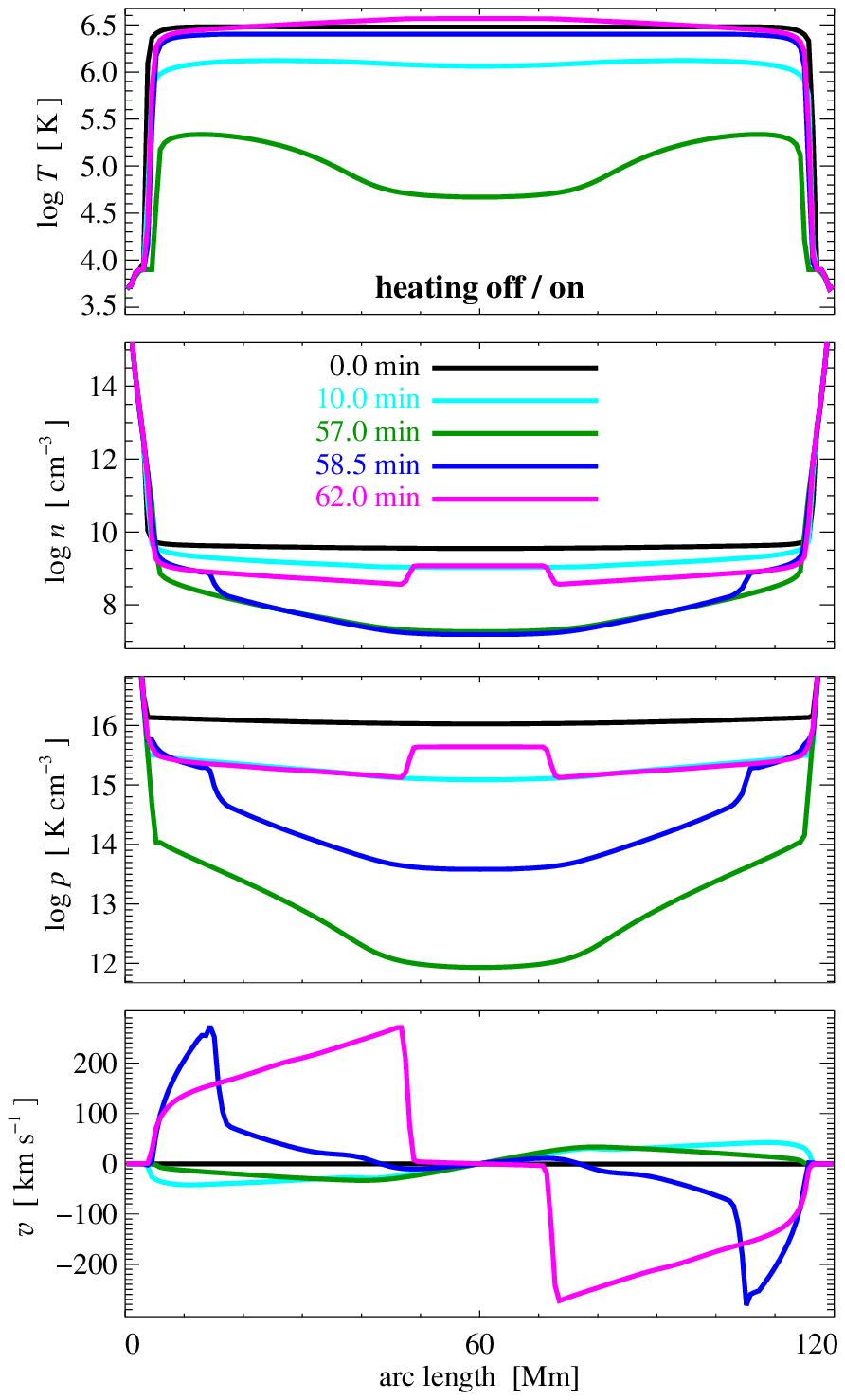} \hfill
\includegraphics[width=\singlefigwid]{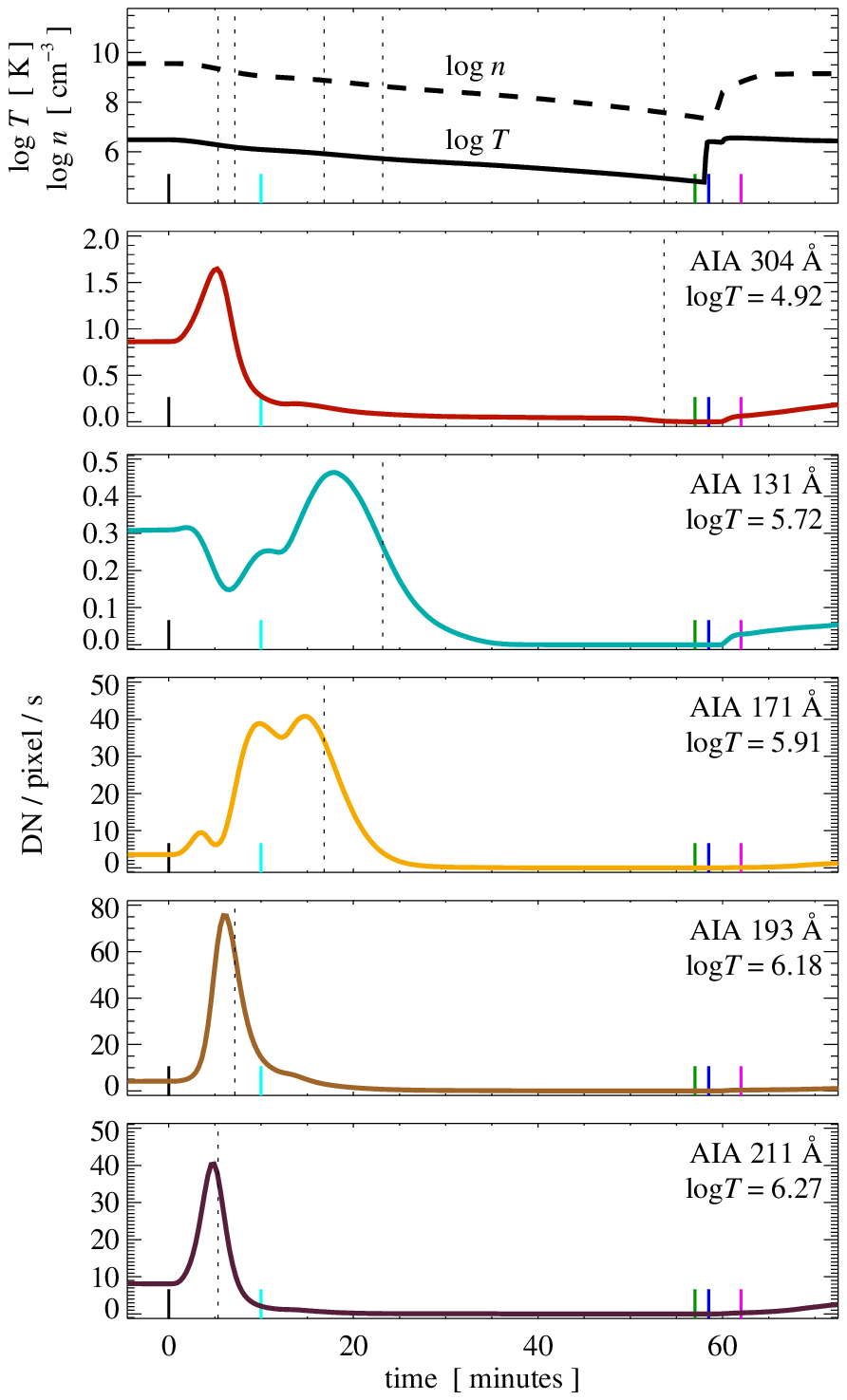}
\caption{Same as \fig{F:condensation} but for a loop with heating switched off at $t{=}0$\,min and on again at ${\approx}57$\,min. See \sects{S:results.cooling} and \ref{S:signature.cooling}.
A movie of the temporal evolution shown in the left panels is available in the on-line edition.
\label{F:cooling}}
\end{figure*}

For the case of shutting off the heating, the temperature and the pressure of the loop respond immediately, which is clearly visible in \fig{F:temporal.global}a,c. The temporal evolution at the top of the loop shown in \fig{F:cooling} (top right) gives a quantitative measure, showing that both temperature and the density fall off roughly exponentially on a time scale of about 10 min, which is comparable to the sound crossing time. Because the energy losses through radiation and heat conduction are no longer replenished, the plasma cools and consequently loses its support: the loop quickly drains.
This is visible in \fig{F:temporal.global}d, which shows the velocity along the loop.

This draining of mass is a pointed difference to the case of catastrophic cooling discussed in the preceding section. Once the condensation has formed there, it goes on to acquire mass, which is visible as upflows in both legs of the loop (\fig{F:temporal.global}h starting at $t{\approx}120$\,min). In contrast, downflows in both legs can be observed in the case with the heating shut off (\fig{F:temporal.global}d). This draining is efficient enough that after some 50 min the density dropped by about a factor of 100, and because of the drop in temperature the pressure at the apex dropped by about $10^4$. This underlines that the case of catastrophic cooling can be considered as a basically isobaric process, at least its pressure change is smaller than in the case of shutting off the heating. This is because when basically following (\ref{E:RTV.T}) and (\ref{E:RTV.n}) the total heating rate sets the pressure, $p{\propto}F_{\rm{H}}^{6/7}$, and the total heat input $F_{\rm{H}}$ remains the same when there is of catastrophic cooling (cf.\ \sect{S:setup.catastrophic}). This has an important consequence when discussing the signatures that are observable with AIA (\sect{S:signature}).

After the loop has cooled and drained we switch the heating on again (at $t{\approx}57$\,min) to return to the value of the equilibrium model. This leads to a very fast increase in the temperature. Within one minute the temperature at the apex rises to the original value of about ${\log}T\,[\mbox{K}]{\approx}6.5$ (cf.\ \fig{F:cooling}). The increased heating rate leads to evaporation at the loop footpoints, and the upflows fill the loop to its former density within some 15\,min (cf.\ \fig{F:cooling}, left panels). This very strong change in density is also pointedly different from the catastrophic cooling, where the density in the loops outside the condensation region remains more or less constant (cf.\ \fig{F:condensation}).

\section{Observational signatures of the cooling events\label{S:signature}}

Because the AIA bandpasses have contributions from multiple temperatures, the expected count rates in a cooling environment do not necessarily show a peak when the plasma reaches the temperature of maximum contribution. This has already been pointed out in our observational study by employing a simplistic model for a cooling plasma at constant pressure \citep[Sect.\,3.3 and Fig.\,6 of][]{Kamio+al:2011.upflow}. The numerical loop model presented here supports and significantly refines our previous arguments.

In the following we analyse the AIA emission, as can be expected at the loop apex when looking horizontally at the loop edge-on (i.e., the observer is in the same plane as the loop). To capture the condensation at the loop top, we concentrate on the part that would be covered by one AIA pixel, i.e.\ ${\approx}0.6\arcsec\,\widehat{=}\,450$\,km. Because we look at the loop apex edge-on, this corresponds to an integration over an arc length of about 6\,Mm to both sides of the apex.
We discuss only the 304\,\AA, 131\,\AA, 171\,\AA, 193\,\AA, and 211\,\AA\ channels of AIA. The 335\,\AA\ channel has a very broad contribution, which is not really useful for studying a condensation, and for the 94\,\AA\ channel it is most likely that the temperature response function based on CHIANTI calculations has to be severely corrected \citep{Aschwanden+Boerner:2011}.

\subsection{Catastrophic cooling\label{S:signature.catastrophic}}

All the AIA channels under investigation show very low count rates at the initial stage of the catastrophic cooling experiment (which is by design, see \sect{S:heating}).
In \fig{F:condensation} (right panel) we show the temporal evolution of the synthesized AIA counts starting at about $t{=}90$\,min, because the early evolution is very slow (cf.\ \sect{S:results.catastrophic}).

\paragraph{``Hot'' channels, 193\,\AA\ and 211\,\AA.}

In these we first see a slow increase in the count rate. This is because the temperature is falling and the broad peak very roughly coincides with the time when the apex reached the temperature, where the temperature response of the respective channel is at maximum (at ${\log}T[\mbox{K}]{\approx}6.2$ and 6.3; indicated by vertical dotted lines). The considerable time lag between reaching the peak of the  temperature response and reaching the peak of the count rate comes about because the pressure is roughly constant, and thus the density increases. This is due to the weighting effect with the density, because the count rate is given through the temperature response weighted by the density squared.

After this broad first peak, a very narrow one can be seen in the 193\,\AA\ and 211\,\AA\ channels (at $t{\approx}120$\,min) when the condensation sets in. It occurs when the apex temperature reaches the value corresponding to the secondary major contribution to these channels at ${\log}T[\mbox{K}]{\approx}5.4$ (cf.\ \fig{F:response}). This second peak is due to ions such as \ion{O}{6} \citep{ODwyer+al:2010}.
It is much sharper in time, because the temperature falls rapidly when forming the condensation. Even though the secondary maximum of the temperature response is a factor of about 10 lower than the main maximum (\fig{F:response}) and the source volume is much smaller (because the transition region is very narrow), the count rate during this second peak is comparable to the first peak (in the case shown in \fig{F:condensation} only a factor of 2 lower). This is because the condensation evolves basically at constant pressure (cf.\ \sect{S:results.catastrophic}): when reaching ${\log}T[\mbox{K}]{\approx}5.4$ the temperature drops by a factor of 10, thus the density increased by a factor of 10, and the temperature response has to be weighted with the density squared!

Finally, once the condensation has slid away from the apex, the plasma resumes its original temperature and density, and we see a third maximum in the 193\,\AA\ and 211\,\AA\ channels, which is basically a mirror of the first maximum. Because the density does not immediately fully return to its original value, the count rates in this second maximum are a bit lower than during the first maximum.

\paragraph{``Cool'' channels, 131\,\AA\ and 171\,\AA.}

These cooler channels, with a maximum contribution at ${\log}T[\mbox{K}]{\approx}5.7$ and 5.9, show basically the same as the ``hot'' channels for the first and the third maximums in the count rate. However, the temporal evolution in the ``cool'' channels is faster; i.e., the peaks are narrower, because the condensation crosses the maximum contribution temperatures  faster (with decreasing temperature the first and the last peaks become increasingly narrow, cf.\ \fig{F:condensation}).

The major difference between the ''hot'' and the ``cool'' channels is that the 131\,\AA\ and 171\,\AA\ channels do not show the second maximum. This is because these channels show a lesser degree of contamination with cooler transition region lines; in particular, they do not show any secondary peak at temperatures below the main peak (even though the contribution shows some extended wing; cf.\ \fig{F:response}).

\paragraph{\ion{He}{2} channel, 304\,\AA.}

The \ion{He}{2} channel shows a strongly asymmetric temporal variation. It shows a first peak when the temperature falls below $10^5$\,K (at $t{\approx}120$\,min). The second broad peak is from the accumulation of cool material near the loop apex as long as the condensation is close to the apex. (This second peak is \emph{not} due to the contribution of hot plasma near ${\log}T[\mbox{K}]{\approx}6.2$ through a \ion{Si}{11} line, cf.\ \fig{F:response}). 
The emission in \fig{F:condensation} is calculated for looking at the loop apex edge-on in one AIA pixel, which corresponds to an integration over a \emph{height} around the apex equivalent to 450\,km, or to an arc length of $\approx$6 Mm to both sides of the apex.
The count rate in this channel drops to very low values again once the condensation starts sliding down the loop and leaves the region of the loop contributing to the the AIA pixel looking at the loop apex from edge-on. As can be seen from \fig{F:temporal.global} (bottom row), the condensation is at an arc length of about 54\,Mm, i.e., some 6\,Mm away from the apex at time $t{\approx}143$\,min . This is also the time when the emission in the 304\,\AA\ channel drops to almost zero.

This 304\,\AA\ light curve has to be taken \emph{cum grano salis}, because the formation of the He lines in the extreme ultraviolet is still not fully understood, which then also applies to the synthesized count rates in the 304\,\AA\ channel.

\subsection{Shut-off heating\label{S:signature.cooling}}

The situation is quite different for the synthesized count rates when the heating is shut off. The cooling sets in, and along with it the draining starts immediately. Thus the density will be considerably lower in this case, leading to significantly lower count rates than in the the catastrophic cooling case.

\paragraph{``Hot'' channels, 193\,\AA\ and 211\,\AA.}

The peak in count rate for these channels slightly precedes the time when the temperature of maximum contribution is reached (see \fig{F:cooling}). This is because the count rate is the temperature response weighted with the density squared and the loop is draining quickly, leading to a rapidly falling density. More importantly, the rapidly falling density also causes the count rates to be very low when shutting off the heating --- considerably lower than for catastrophic cooling.

After 5\,min, when the peak in 211\,\AA\ is reached, the density has \emph{dropped} already by a factor of 2. In contrast, during the first peak in 211\,\AA\ for the catastrophic cooling ($t{\approx}105$\,min, \fig{F:condensation}), the density has \emph{increased} by some 60\% to 70\%. Thus at the time of the (first) 211\,\AA\ peak the density in the catastrophic cooling case is higher by a factor of slightly more than 3 as compared to the case of shutting off the heating. This leads to the difference by a factor of about 10 in 211\,\AA\ count rate during the first 211\,\AA\ peak (compare lower right-hand panels of \figs{F:condensation} and \ref{F:cooling}). The same arguments hold for the 193\,\AA\ channel, which shows an even larger difference because it peaks later.

Because the density drops so rapidly, the secondary maximum of the temperature response function at low temperatures does not play a role here. Thus, unlike the catastrophic cooling there there is only a single peak in each of these ``hot'' channels.

\paragraph{``Cool'' channels, 131\,\AA\ and 171\,\AA.}

For the same reason as the ``hot'' channels these ``cool'' ones show much lower count rates than in the case of catastrophic cooling, only now the difference is even greater, up to a factor 1000, because they peak later when even more material has drained. At this very low count rate level, they show a more complex light curve, because of the multi-peak structure of the temperature response at higher temperatures (cf.\ \fig{F:response}).

\paragraph{He\,{\small{II}} channel, 304\,\AA.}

When employing the AIA temperature response, this channel shows only one single peak early on. This is because of the side maximum of the 304\,\AA\ temperature response near ${\log}T[\mbox{K}]{\approx}6.2$. Consequently this peak appears almost simultaneously with the 193\,\AA\ channel peak. Later in the evolution of the cooling loop, the density is simply too low to produce a noticeable signal in this channel.

\paragraph{Switching on the heating again.}

After switching on the heating rate again at $t{\approx}57$\,min, the temperature almost immediately resumes its original temperature (\sect{S:results.cooling}). During the subsequent filling of the loop with material through evaporation, the count rates in all channels creep up \emph{monotonically} until they reached the initial values at $t{=}0$ (not fully shown in \fig{F:cooling}). In particular, no other peaks are found in the count rates.

\begin{figure}
\includegraphics[width=\singlefigwid]{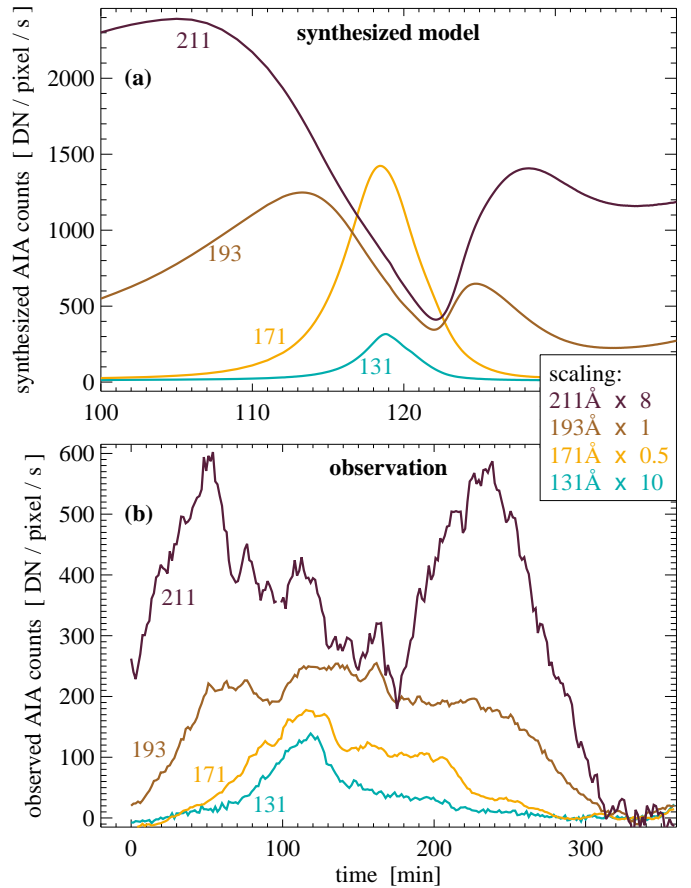}
\caption{Comparison of synthesized (top) and observed (bottom) emission of catastrophic cooling in SDO/AIA bands. The count rates for the AIA bands are scaled using the factors listed in the top panel in order to fit better in a single plot. Both synthesized and observed rates are scaled by the \emph{same} factors. The synthesized count rates are for an edge-on observation of the loop near the apex averaged over 1.3\,Mm (${\approx}3$ AIA pixels).
See \sect{S:obs}.
\label{F:comparison}}
\end{figure}

\section{Comparison to observations\label{S:obs}}

One conclusion from the discussion in \sect{S:signature} is that a simple shut-down of the heating rate cannot produce significant observable signatures above the limb that would be detectable with AIA/SDO, simply because the expected count rates would be too low. Even if one increased the density of the initial loop (to unrealistically higher values), the quick draining would prevent significant count rates. This rules out that simply shutting off the heating rate could reproduce the cooling structures seen in condensations above the limb.

Catastrophic cooling provides a mechanism that produces high count rates in the AIA bands \emph{and} gives complex (not single-peaked) light curves. In the following we compare the synthesized count rates of the catastrophic cooling model to a recent AIA observation by \cite{Kamio+al:2011.upflow} of a condensation above the limb.

The bottom panel of \fig{F:comparison} shows the light curves in a single AIA pixel of the same structure above the limb as already discussed by \cite{Kamio+al:2011.upflow} and shown in their Fig.\,5. While \cite{Kamio+al:2011.upflow} show the normalized profile, we now give the actual count rates (with some scaling so that all curves fit into the same panel). To subtract the background contribution, we corrected for the count rates found after the condensation event (in our plot the average after time $t{\approx}320$\,min).
The light curves clearly reveal single peaks in the ``cool'' 131\,\AA\ and 171\,\AA\ channels. The ``hot'' channels show multi-peaked structures, with 193\,\AA\ having a broad maximum with maybe two or three peaks, and 211\,\AA\  a clear double peak and some indication of a third peak in-between, co-temporal with the 131\,\AA\ and 171\,\AA\ peaks.

In the top panel of \fig{F:comparison} we show the count rate synthesized from the model looking horizontally at the apex edge on. This is similar to \fig{F:condensation}, but now we have averaged over a region corresponding to three AIA pixels in height (equivalent to 1.3\,Mm). This is done to get a better match to the actual AIA resolution, which is (predicted to be) of the order of 1.6\arcsec\ corresponding to roughly three pixels in the channels used in our study \citep[Table 8 of][]{Boerner+al:2011}. As for the observations in the bottom panel, we employ a scaling so that all curves fit into the same panel. The scaling for the data synthesized from the model and the actual observation is \emph{identical}.

Because of the averaging, the second peak in the ``hot'' channels near $t{=}120$\,min is not visible here. This very narrow peak is strongly located in space at the place where the condensation occurs and is thus outshone by the emission along the top part of the loop we average over when looking at the top 1.3 Mm at the loop apex (corresponding to an arc length of more than 10\,Mm). The structure that remains in the light curves synthesized from the model is a broad double peak in the ``hot'' channels 193\,\AA\ and 211\,\AA\ and a narrower single peak in the middle in the ``cool'' channels 131\,\AA\ and 171\,\AA.

Comparing the data synthesized from the model and the actual observations in \fig{F:comparison}, it is obvious that some general features match. The ``hot'' channels show a broad double-peak structure, the single peaks in the ``cool'' channels are comparably narrow and appear roughly in the middle of the ``hot'' channel light curves, and finally the order of magnitude of the predicted count rate matches the observations (within a factor of about 5) and the ratios of the different channels roughly match (within a factor of 2). 
While we do not discuss the \ion{He}{2} 304\,\AA\ channel further because of the problems with our knowledge of the He line formation, it is noticeable that the observation by \cite{Kamio+al:2011.upflow}, their Fig.\,5, shows a narrow peak in 304\,\AA\ (co-temporal with the 171\,\AA\ peak) followed by a broad second peak. This is similar to our synthesized light curve for 304\,\AA\ in \fig{F:condensation}.

All this is achieved without fine-tuning the model; i.e., we did \emph{not} run a large number of models and just picked the best possible match. As outlined in \sect{S:heating} the only real free parameters are the loop length $L$, the heat input $F_{\rm{H}}$, and the scale length of the heating rate $\lambda$. However, for the comparison these are not free to choose: $L$ is set by the height above the limb where the condensation is observed, and $F_{\rm{H}}$ is set by the requirement to reach a apex temperature above the formation temperature of the 211\,\AA\ channel (and we chose ${\log}T[\mbox{K}]{\approx}6.5$). This then also sets the pressure and the density in the loop. For $\lambda$ we had to choose a value so that the catastrophic cooling occurs. In the end, there is not that much room to play with the model parameters.

Of course, there is also a significant difference between a model and observations. In the model the enhancement in the synthesized AIA light curves is restricted to some 20\,min in the ``cool'' channels and some 60\,min in the ``hot'' channels. The corresponding time scales in the observations are about a factor of five longer. Because the time span of the condensation depends on the the heating scale length $\lambda$ \citep[e.g.][]{Mueller+al:2003,Mueller+al:2004}, an alteration of this parameter could lead to a better match. Also including of non-equilibrium ionization could work into this direction, as it will lead to longer time scales for the condensations \citep{Mueller+al:2004}.

\section{Conclusions\label{S:conclusions}}

We ran two types of models for cooling in coronal loops and compared them to observations. In the \emph{heating off/on} model we simply shut down the heating, which leads to cooling and draining of the loop, and then resumed the original heating. In the \emph{catastrophic cooling} model we kept the total heat input constant but concentrate the heating more towards the footpoints. This leads to a thermal non-equilibrium near the apex, and a condensation forms in a runaway process. While these two processes have been studied in the past, we put them here in the new context of observations at the solar limb in extreme ultraviolet passbands by AIA/SDO.

We found that the \emph{heating off/on} model will not generate significant count rates to understand observed condensation events above the limb. In contrast, the \emph{catastrophic cooling} model seems to provide a natural explanation for the observations of condensations, with a good match to the observed light curves. Not only are the observed count rates roughly matched, but also the single peaked nature of the lightcurves in ``cool'' channels (131\,\AA\ and 171\,\AA), and the multiple peaks in the ``hot'' channels (193\,\AA\ and 211\,\AA) are reproduced.

Good arguments have been presented that the process of thermal non-equilibrium cannot be used as a general process to understand all properties of coronal loops \citep{Klimchuk+al:2010}. Nonetheless, this process is important in a subvolume of the corona, e.g.\ where the coronal condensations form, and it is a valuable tool for investigating the dynamic response of the corona to the spatial distribution of the heat input into the corona.

The important conclusion from our study is that, even in condensation events in the corona, a constant supply of energy is needed to keep the coronal pressure. The \emph{heating off/on} model fails, because the support of the corona is lost when shutting down the heating and the material drains very quickly. To see condensations in the corona, one has to keep the energy input at a sufficient magnitude to support the hot plasma further. In the  \emph{catastrophic cooling} model, this is achieved by a higher concentration of the heating towards the loop footpoints, which reduces the heat input in the \emph{high} corona but keeps the \emph{total} amount of energy supply to the corona at a (more or less) constant level.

\acknowledgements
The data used are provided courtesy of NASA/SDO and the AIA science team. The authors are grateful for the constructive comments from the referee. Special thanks are due to Zoran Miki\'c for our discussions on the process of thermal non-equilibrium.



\end{document}